\documentclass[%
 reprint,
 aps,
 pra,
 superscriptaddress
]{revtex4-2} 

\usepackage{amssymb,amsthm,amsmath,amsfonts}
\usepackage{graphicx}
\usepackage[caption=false]{subfig} 
\usepackage{bm} 
\usepackage{bbm}
\usepackage{dcolumn} 
\usepackage{xcolor}
\usepackage{hyperref}
\hypersetup{
    colorlinks,
    linkcolor={red!50!black},
    citecolor={blue!50!black},
    urlcolor={blue!80!black}
}
\usepackage{tikz}
    \usetikzlibrary{shapes, arrows.meta, positioning}

\usepackage{layouts} 

\usepackage[nameinlink, capitalize]{cleveref}
\crefname{section}{Sec.}{Secs.}
\Crefname{section}{Section}{Sections}

\graphicspath{ {./pics/} }

\newcommand{\id}{\mathbbm{1}} 
\DeclareMathOperator{\spann}{span}
\newcommand{\rmi}{\mathrm{i}}
\newcommand{\rme}{\mathrm{e}}

\newcommand{\linewidthfull}{16.3cm}
\newcommand{\linewidthhalf}{8.6cm}

\begin{document}

\title{Multipartite entanglement dynamics in quantum walks}

\author{Emil K. F. Donkersloot}
\email{emil.donkersloot@uni-jena.de}
\affiliation{Institute of Condensed Matter Theory and Optics, Friedrich-Schiller-University Jena, Max-Wien-Platz 1, 07743 Jena, Germany}

\author{Ren\'e Sondenheimer}
\affiliation{Institute of Condensed Matter Theory and Optics, Friedrich-Schiller-University Jena, Max-Wien-Platz 1, 07743 Jena, Germany}
\affiliation{Fraunhofer Institute for Applied Optics and Precision Engineering IOF, Albert-Einstein-Stra\ss{}e 7, 07745 Jena, Germany}

\author{Jan Sperling}
\email{jan.sperling@uni-paderborn.de}
\affiliation{Theoretical Quantum Science, Institute for Photonic Quantum Systems (PhoQS),
Paderborn University, Warburger Stra\ss{}e 100, 33098 Paderborn, Germany}

\begin{abstract}

    Quantum walks constitute a rich area of quantum information science, where multipartite entanglement plays a central role in the dynamics and scalability of quantum advantage over classical simulators.
    In this work, we study the multipartite entanglement of quantum walks in optical settings.
    We present methods for computing a geometric measure of entanglement for arbitrary partitions of a single-walker quantum walk and for analyzing the entanglement in multi-walker scenarios.
    These techniques are used for numerical studies on the entanglement dynamics of quantum walks in large systems and under various initial conditions.
    For a given bipartition, based on the coin degrees of freedom, we derive exact expressions describing the complete entanglement dynamics for arbitrary localized initial conditions.
    We use these expressions for analytic statements about the asymptotic behavior of the system.
    Furthermore, we demonstrate the emergence of entanglement typicality in statistical ensembles of random optical networks.

\end{abstract}


\maketitle

\section{Introduction}\label{sec:Introduction}

    Quantum walks (QWs) were introduced \cite{aharonovQuantumRandomWalks1993} as a quantum version of the classical random walk.
    Since then, QWs have gone far beyond a quantum analog of a classical walk, constituting a contemporary area of research in quantum information science \cite{kadianQuantumWalkIts2021,zhouMultiparticleQuantumWalks2024,difidioQuantumWalksEntanglement2024}.
    A main practical hallmark of QWs is that they were recognized early as a suitable system for photonic implementations \cite{bouwmeesterOpticalGaltonBoard1999,zhaoImplementQuantumRandom2002}, which was further extended to other platforms \cite{venegas-andracaQuantumWalksComprehensive2012}.
    One way to implement QWs in a photonic system uses linear optical networks (LONs) \cite{zhaoImplementQuantumRandom2002}, which have gained increasing attention because of their prominent role in quantum computing \cite{knillSchemeEfficientQuantum2001} and relevance for the seminal boson sampling protocol \cite{aaronsonComputationalComplexityLinear2011,hamiltonGaussianBosonSampling2017}.

    Entanglement plays an essential role in the function of a QW;
    see, e.g., \cite{horodeckiQuantumEntanglement2009} for an overview.
    Although entanglement characterization is paramount for understanding such dynamical quantum systems, its computation presents a major technical challenge \cite{gurvitsClassicalDeterministicComplexity2003}.
    Thus, there have been many pioneering studies on entanglement in LONs and QWs \cite{carneiroEntanglementCoinedQuantum2005,abalQuantumWalkLine2006,ideEntanglementDiscretetimeQuantum2011,rohdeEntanglementDynamicsQuasiperiodicity2012,ortheyAsymptoticEntanglementQuantum2017,panMultiphotonEntanglementInterferometry2012,qiaoEntanglementFullState2023,iosuePageCurvesTypical2023}.
    However, most have focused on computing bipartite entanglement, either directly or by considering multipartite entanglement measures which rely on bipartitions.
    Bipartite entanglement is generally more approachable, but it cannot capture the full entanglement structure of large multipartite systems \cite{gerkeMultipartiteEntanglementTwoSeparable2016}.
    Still, finding suitable measures for multipartite entanglement, which do not rely on bipartitions, remains a challenging problem.
    One such measure is the geometric measure of entanglement \cite{weiGeometricMeasureEntanglement2003}, which indicates how far a quantum state is from the set of all separable (i.e., non-entangled) states.
    We can utilize the geometric measure of entanglement to construct optimized entanglement witnesses \cite{weiGeometricMeasureEntanglement2003,sperlingMultipartiteEntanglementWitnesses2013}, which constitute a valuable tool in experiments, even under the influence of noise;
    see, e.g., Ref.\ \cite{barbieriDetectionEntanglementPolarized2003}.
    Here, we focus on the calculation of the geometric measure of entanglement of QWs in an optical setting.

    One of the early studies on entanglement in QWs was reported in Ref.\ \cite{carneiroEntanglementCoinedQuantum2005}.
    There, the von Neumann entropy \cite{nielsenQuantumComputationQuantum2012} between the coin and the positions of a walker was studied numerically for various settings.
    For the asymptotic case, some numerical results in Ref.\ \cite{carneiroEntanglementCoinedQuantum2005} were later proven in Refs.\ \cite{abalQuantumWalkLine2006, ideEntanglementDiscretetimeQuantum2011}, which is important to understand the long-term behavior of quantum effects.
    In Ref.\ \cite{rohdeEntanglementDynamicsQuasiperiodicity2012}, the multipartite entanglement dynamics for a QW of finite size, including reflective boundary conditions, was studied using the Meyer--Wallach entanglement measure \cite{meyerGlobalEntanglementMultiparticle2002}; however, this measure is based on bipartitions.
    For small systems, they showed the quasi-periodic behavior of the entanglement dynamics, stemming from the finite size of the system under study.

    In this work, we study the multipartite entanglement in QWs for finite and infinite systems.
    We develop an efficient approach to compute the multipartite geometric measure of entanglement for QWs with a single walker.
    This method is applicable to all possible partitions of the full composite system.
    We study how the multipartite entanglement evolves in quantum walks of large sizes and how the initial conditions and chosen partitions influence the dynamics of entanglement.
    Additionally, we derive analytical results for a bipartition of a QW on the line for arbitrary time steps, allowing us to investigate the asymptotic entanglement. 
    As an extension, we study the genuine multipartite entanglement for smaller systems in the case of multiple walkers.
    Our study reveals distinct temporal behaviors of entanglement in QWs and enables the exploration of entanglement typicality in random networks.

    The paper is structured as follows.
    In \cref{sec:QuantumWalks,sec:Entanglement}, we introduce the general framework.
    In \cref{sec:FullSepOnePhoton}, we present the method we use to efficiently compute the entanglement of systems with a single walker.
    We start applying this method by considering statistical ensembles of QWs, i.e., LONs, in \cref{sec:Random}.
    We then derive analytical results for the entanglement of a single walker on the line for arbitrary time steps in \cref{sec:CoinEntLine} and consider the evolution of entanglement for a QW on a circle in \cref{sec:EntanglementCircle}.
    We close with an instructive outlook on the case of multiple walkers in \cref{sec:MultiplePhotons} by studying the geometric measure of entanglement and genuine multipartite entanglement.
    Finally, we conclude in \cref{sec:Conclusion}.

\section{Methods}\label{sec:Methods}
    
    We begin by briefly summarizing our notation and conventions used throughout this paper.
    We consider networks with $M$ quantized modes and $N$ photons, likewise walkers.
    A tensor-product state in the composite Hilbert space $\mathcal{H} = \mathcal{H}_1 \otimes \mathcal{H}_2$ can be denoted by $|\chi_1\rangle \otimes |\chi_2\rangle \equiv |\chi_1\rangle|\chi_2\rangle \equiv |\chi_1 \chi_2\rangle \equiv |\vec{\chi}\rangle$, where $|\chi_1 \chi_2\rangle \in \mathcal{H}$ and $|\chi_i\rangle \in \mathcal{H}_i$.
    An LON, represented via the unitary $U \in \mathrm{U}(M)$, transforms an input state $|\psi_0\rangle$ into the output state $|\Psi\rangle=\hat\varphi(U)|\psi_0\rangle$, where $\hat\varphi(U)$ is the unitary representation from the mode picture to the state space, $\hat\varphi: \mathrm{U}(M) \to \mathrm{U}(\dim \mathcal{H})$ \cite{aaronsonComputationalComplexityLinear2011}.
    A single time step is denoted by $\hat T\in\mathrm{U}(\dim \mathcal{H})$;
    after $n \in \mathbb{N}$ time steps, the walk evolves to $|\Psi(n)\rangle = \hat T^n |\psi_0\rangle$, where $|\Psi(0)\rangle = |\psi_0\rangle$.
    For a two dimensional coin, the QW corresponds to an LON with $M = 2P$ modes, where $P$ is the number of possible positions of the QW.
    The geometric measure of entanglement is denoted by $E_g$.
    When applicable, we use the common multi-index notation for convenience.
    A multi-index is denoted by $\vec n \in \mathbb{N}^M$ with components $(\vec n)_i = n_i$.
    A quantity $\psi$ with $M$ indices is then written as $\psi_{\vec n} = \psi_{n_1 n_2 \ldots n_M}$, and photon-number states are denoted by $|\vec n\rangle = |n_1,n_2,\ldots,n_M\rangle$.
    The sum of the indices is written as $|\vec n| = \sum_{i=1}^M n_i$, which corresponds to the total number of photons.

    \subsection{Quantum walks and optical networks} \label{sec:QuantumWalks}
    
        \subsubsection{Quantum walks}
        
            The idea of a quantum walk is that the Hilbert space $\mathcal{H}$ is split into a coin space $\mathcal{H}_C$ and a position space $\mathcal{H}_P$; see  \cite{kempeQuantumRandomWalks2003,portugalQuantumWalksSearch2018,venegas-andracaQuantumWalksComprehensive2012,kadianQuantumWalkIts2021} for reviews.
            The time evolution is then implemented by first performing a coin flip $\hat C : \mathcal{H}_C \to \mathcal{H}_C$, followed by a controlled step operation $\hat S: \mathcal{H}_C \otimes \mathcal{H}_P \to \mathcal{H}_C \otimes\mathcal{H}_P$, where the coin acts as the control.
            The example that is mostly studied in this work is the Hadamard walk \cite{nayakQuantumWalkLine2000}, where the coin space is two dimensional $\mathcal{H}_C = \spann\{ |-\rangle, |+\rangle \}$ and the coin operation is given by the Hadamard gate
            \begin{equation}
                \hat C = \hat H = \frac{1}{\sqrt{2}} \begin{pmatrix}
                    1 & 1 \\
                    1 & -1
                \end{pmatrix} \; \Rightarrow \; \hat H|\mp\rangle = \frac{|-\rangle \pm |+\rangle}{\sqrt{2}}\,.
            \end{equation}
            The states $|\pm\rangle$ should not be confused with the eigenstates of the Pauli-$X$ operator. 
            For example, our $|-\rangle$ would correspond to $|0\rangle$ in the typical quantum information notation.
            The notation $|\pm\rangle$ is chosen to indicate the walker's step direction.
            The position space is given by $\mathcal{H}_P = \mathrm{span} \{ |x\rangle \,|\, x\in\mathbb{Z} \}$, with the step operation
            \begin{align}\label{eq:StepOperator}
                \hat S &= | - \rangle\langle - | \otimes \sum_{x \in \mathbb{Z}} |x-1\rangle\langle x | \nonumber
                \\
                &\quad +\, | + \rangle\langle + | \otimes \sum_{x \in \mathbb{Z}} |x+1\rangle\langle x |\, .
            \end{align}
            That is, if the coin is in the `$-$' state, the walker moves one step left, $x \mapsto x-1$, and if the coin is in the `$+$' state, the walker moves one step right, $x \mapsto x+1$. 
            With infinite positions ($x \in \mathbb{Z}$), we refer to the QW as a quantum walk `on the line'.
            One can also restrict oneself to a finite number of positions $x \in \{0,\ldots,P-1\}$, where $P \in \mathbb{N}$.
            In this case, we have to define what happens at the boundaries $x = 0$ and $x = P-1$.
            We choose periodic boundaries: $|x = -1\rangle \equiv |x = P-1\rangle$ and $|x = P\rangle = |x = 0\rangle$, resulting in a QW `on a circle'.
            
            When starting with the initial state $|\Psi(0)\rangle \equiv |\psi_0\rangle$, a single time step is given by $|\psi_0\rangle \mapsto \hat T|\psi_0\rangle$, where $\hat T = \hat S (\hat C \otimes \id)$.
            Evolving the state $n \in \mathbb{N}$ time steps gives the state $|\Psi(n)\rangle \equiv \hat T^n|\psi_0\rangle$.

        \subsubsection{Linear optical networks}\label{sec:LONs}

            LONs can be described via the ladder-operator formalism \cite{vogelQuantumOptics2006}.
            The $M$ input modes are described by $\hat{a}_i$, with $[\hat a_i, \hat a_j^\dag] = \delta_{ij}$. 
            The $M$ output modes are described by $\hat b_i$, with $[\hat b_i, \hat b_j^\dag] = \delta_{ij}$.
            The propagation of the modes in the LON is given by a unitary matrix $U: \mathbb C^M\to\mathbb C^M$,
            \begin{equation}\label{eq:linearTrafoLadderOp}
                \vec{\hat{a}}^\dag \mapsto \vec{\hat{b}}^\dag = U \vec{\hat{a}}^\dag,
            \end{equation}
            where $\vec{\hat{a}}^\dag = (\hat a^\dag_i)_{i=1}^M$ and $\vec{\hat b}^\dag=(\hat b^\dag_i)_{i=1}^M$.
            The effect of the LON on a quantum state can be described by considering photon-number basis states as inputs to the LON,
            \begin{equation}
                \label{eq:InputStatePhotonBasis}
                |\text{in}\rangle = |n_1,\ldots, n_M\rangle \equiv \frac{\hat a_1^{\dag n_1}}{\sqrt{n_1!}} \cdots \frac{\hat a_M^{\dag n_M}}{\sqrt{n_M!}} |\text{vac}\rangle.
            \end{equation}
            The output state $|\text{out}\rangle \equiv |\psi\rangle = \hat\varphi(U)|\text{in}\rangle$ is given by updating $\hat a_i^\dag$ with $\hat b_i^\dag$, as prescribed by \cref{eq:linearTrafoLadderOp}.
            The operator $\hat\varphi(U):\mathcal H\to\mathcal H$ denotes the unitary representation in state space associated with the unitary evolution $U$ in the mode picture. 
            The coefficients of $|\text{out}\rangle$ can be given by permanents of matrices constructed from $U$ \cite{scheelPermanentsLinearOptical2004}.
            Note that the computational complexity of permanents yields the quantum advantage of the boson sampler \cite{aaronsonComputationalComplexityLinear2011}.
            
            An important consequence of an LON (without loss) is photon-number conservation: if $N$ photons enter the network, then $N$ photons exit the network.
            The state $|\psi_0\rangle = |\vec n\rangle$ has $N = |\vec n|$ photons distributed across all modes.
            The output state $|\psi\rangle = \hat\varphi(U)|\psi_0\rangle$ is thus of the following, general form: 
            \begin{equation}\label{eq:OutputLON}
                |\psi\rangle = \sum_{\vec n:\, |\vec n|=N} \psi_{\vec n} |\vec n\rangle.
            \end{equation}

            One can now connect, for example, the Hadamard walk, described above, with an LON, consisting of $M = 2P$ modes.
            We can then label the modes $i=0,\ldots,M-1$ in the following way:
            \begin{equation}
                i = P c + x,
            \end{equation}
            where $x\in\{0,\ldots,P-1\}$ and $c\in\{ 0,1\}$.
            Here, $c=0$ corresponds to the coin state `$-$' and $c=1$ corresponds to `$+$'.
            To make the decomposition even clearer, we can write $\vec{\hat{a}}^\dag = (\vec{\hat{a}}^\dag_-, \vec{\hat{a}}^\dag_+)$, where $\vec{\hat{a}}^\dag_- = ({\hat{a}}^\dag_0,\ldots,{\hat{a}}^\dag_{P-1})$ and $\vec{\hat{a}}^\dag_+ = ({\hat{a}}^\dag_{P},\ldots, {\hat{a}}^\dag_{M-1})$.
            Thus, the Hadamard walk can be given by the block matrix
            \begin{equation}
                U = \frac{1}{\sqrt{2}} \begin{pmatrix}
                    \Sigma & \Sigma \\
                    \Sigma^\mathsf{T} & -\Sigma^\mathsf{T}
                \end{pmatrix},
            \end{equation}
            where $\Sigma$ is the translation matrix with periodic boundary conditions
            \begin{equation}
                \Sigma = \begin{pmatrix}
                    0 & 1 & 0 & \cdots & 0 \\
                    0 & 0 & 1 & & 0 \\
                    \vdots & & & \ddots & \vdots \\
                    0 & 0 & 0 & & 1 \\
                    1 & 0 & 0 & \cdots & 0
                \end{pmatrix},
            \end{equation}
            or $\Sigma_{ij} = \delta_{i=j+1\, \mathrm{mod}\, P}$ in index notation.
            In the case of a walker on the line, one selects $P$ sufficiently large such that the time steps are insufficient to reach the boundary.

    \subsection{Entanglement} \label{sec:Entanglement}

        Next, we briefly review the concepts of entanglement, entanglement witnesses, and the geometric measure of entanglement, together with their formulation in terms of the so-called separability eigenvalue equations.

        A pure quantum state $|\chi\rangle \in \mathcal{H}$ is separable with respect to a given partition $\mathcal{P}$ of $\mathcal{H} = \mathcal{H}_1 \otimes \cdots \otimes \mathcal{H}_K$ if it can be written as a product state,
        \begin{equation}
            |\chi\rangle \equiv |\vec{\chi}\rangle \equiv |\chi_1 \ldots \chi_K\rangle \in \mathcal{S}_{\mathcal{P}},
        \end{equation}
        where $K \leq M$, $|\chi_i\rangle\in\mathcal H_i$, and $\mathcal{S}_{\mathcal{P}}$ is the set of all separable states regarding the partition $\mathcal{P}$.
        If the state cannot be written in this form, it is said to be entangled.
        We are particularly interested in the modal entanglement of the quantum states that exit an LON. 
        Furthermore, a partition of $\mathcal{P}$ refers to the way this Hilbert space can be decomposed;
        that is, the mode-index set $\mathcal{I} = \{1,\ldots,M\}$ is decomposed into $K$ non-empty and disjoint sets, written as $\mathcal{P} = \mathcal{I}_1:\cdots:\mathcal{I}_K$, where $\mathcal{I}_1 \cup \cdots \cup \mathcal{I}_K = \mathcal{I}$, $\mathcal{I}_k \cap \mathcal{I}_l = \emptyset$ for $k \neq l$, and $\mathcal I_k\neq\emptyset$.
        
        To identify entanglement, one can use entanglement witnesses \cite{guhneEntanglementDetection2009}, providing a necessary and sufficient set of entanglement criteria.
        An entanglement witness is an observable $\hat W$, which satisfies
        \begin{align}
            \nonumber
            \langle \hat W \rangle_{\hat\sigma} &\geq 0 \ \text{for all separable states}\ \hat\sigma \in \mathcal{S}_{\mathcal{P}},
            \\
            \langle \hat W \rangle_{\hat\rho} &< 0 \ \text{for at least one (entangled) state}\ \hat\rho \notin \mathcal{S}_{\mathcal{P}}.
        \end{align}
        An optimal entanglement witness can be written in the following form \cite{sperlingMultipartiteEntanglementWitnesses2013}:
        \begin{equation}\label{eq:EntanglementWitnessDefinition}
            \hat W = g_\text{max}(\hat L) \id - \hat L,
        \end{equation}
        with the Hermitian operator $\hat L$ (also called the test-operator) and $g_\text{max}(\hat L) \in \mathbb{R}$, defined as 
        \begin{equation}
            g_\text{max}(\hat L) = \sup_{\substack{|\vec{\chi}\rangle \in \mathcal{S}_{\mathcal{P}} \\ \langle \vec{\chi}|\vec{\chi}\rangle = 1}} \langle \vec{\chi}|\, \hat L\,| \vec{\chi}\rangle.
        \end{equation}
        In this paper, we use $\hat L = |\psi\rangle\langle\psi|$ for the pure states under study as it always witnesses $|\psi\rangle$ if it is entangled because $g_{\max}\equiv g_{\max}(|\psi\rangle\langle\psi|)=1$ iff $|\psi\rangle=|\vec \chi\rangle$.
        In this case, $g_\text{max}$ is the maximum overlap between the state $|\psi\rangle$ and all separable states.
        Thus, $g_\text{max}$ can be used to construct a geometric measure for how far $|\psi\rangle$ is from the separable states \cite{weiGeometricMeasureEntanglement2003}, which commonly takes the form
        \begin{equation}
            E_g = 1-g_\text{max}.
        \end{equation}
        
        The quantity $g_\text{max}$ can be calculated with the separability eigenvalue (SEV) equations \cite{sperlingMultipartiteEntanglementWitnesses2013}.
        These consist of the following set of coupled eigenvalue equations
        \begin{equation}
            \hat L_{j} |\chi_j\rangle = g |\chi_j\rangle,
        \end{equation}
        where $j\in\{1,\ldots,K\}$.
        Here, $|\chi_j\rangle \in \mathcal{H}_j$ are the normalized eigenstates of the reduced operator 
        \begin{align}\label{eq:SEVGeneral}
            \hat L_{j} ={}& \big[ \langle \chi_1\ldots \chi_{j-1}| \otimes \id \otimes \langle \chi_{j+1}\ldots \chi_K|\big]\, \hat L \, 
            \\ \nonumber
            {}&
            \times \big[ | \chi_1\ldots \chi_{j-1}\rangle \otimes \id \otimes | \chi_{j+1}\ldots \chi_K\rangle \big],
        \end{align}
        and $g$ are the corresponding eigenvalues, given by
        \begin{equation}
            g = \langle \vec \chi |\, \hat L \,|\vec \chi \rangle
            \quad\Rightarrow\quad
            g_\text{max}(\hat L) = \max \{g\},
        \end{equation}
        which is called the separability eigenvalue with the corresponding separability eigenvector $|\vec \chi \rangle$.

    \subsection{Multipartite entanglement for a single-photon walker} \label{sec:FullSepOnePhoton}

        For our first main result, we focus on the case of a single walker ($N=1$) and derive an efficient scheme to compute the multipartite geometric measure of entanglement for the full separability problem, $\mathcal{P} = \{1\}:\cdots:\{M\}$.
        We then show how this method can be easily adapted for arbitrary partitions $\mathcal{P}$.

        \subsubsection{Full partition}\label{sec:EntanglementWState}
        
        Single-photon states are described by generalized (aka asymmetric) W states,
        \begin{equation}\label{eq:WState}
            |\psi\rangle = \sum_{i=1}^M \lambda_i \underbrace{|0_1 \ldots 0_{i-1} 1_i 0_{i+1} \ldots 0_M\rangle}_{\equiv |1_i\rangle}. 
        \end{equation}
        Without loss of generality, the parameters can be chosen to be positive real numbers $\lambda_i \in [0,1]$ that are ordered $\lambda_1 \leq \lambda_2 \leq \ldots \leq \lambda_M$.
        This can always be achieved by local phase shifts and mode reordering, which do not change the entanglement.
        In Ref.\ \cite{tamaryanDualityGeometricMeasure2010}, a method for solving \cref{eq:SEVGeneral} for generalized W states [\cref{eq:WState}] was introduced.
        We adopt and modify this method to be well-suited for numerical computations.
        
        The general idea is to recognize that $g^{-1/2} \prod_{i=1}^M \langle0|\chi_i\rangle$ is an invariant quantity with respect to $|\vec{\chi}\rangle$.
        The problem then reduces to finding this invariant, which further reduces to finding roots of the relatively simple, real-valued functions
        \begin{equation}
            F_{1/2}(\xi) = - \sum_{i=1}^{M-1} \sqrt{1 - \lambda_i^2 \xi} \mp \sqrt{1 - \lambda_M^2 \xi} + M - 2,
        \end{equation}
        where $F_1$ corresponds to `$-$' and $F_2$ corresponds to `$+$'.
        We further define
        \begin{multline}
            g_{1/2}(\xi) = \frac{1}{2^{M-2} \xi} \left[ \prod_{i=1}^{M-1} \left(1 + \sqrt{1-\lambda_i^2 \xi}\right) \right] 
            \\
            \times \left( 1 \pm \sqrt{1 - \lambda_M^2 \xi} \right),
        \end{multline}
        where $g_1$ corresponds to `$+$' and $g_2$ corresponds to `$-$'.
        The following description shows how to use these two functions to solve \cref{eq:SEVGeneral}.
        A proof for this can be found in \cite{tamaryanDualityGeometricMeasure2010}.
        
        There are exactly either $M$ or $M+1$ solutions to \cref{eq:SEVGeneral} for the single-photon states in \cref{eq:WState}.
        The $M$ solutions $g = \lambda_i^2$ for $i=1,\ldots,M$ always exist.
        If $\lambda_M^2 \geq \frac{1}{2}$, then these are the only solutions and $g_\text{max} = \lambda_M^2$.
        If, however, $\lambda_M < \frac{1}{2}$, there is one additional solution, which coincides with $g_\text{max}$.
        To find this solution, one has to solve either $F_1(\xi) = 0$ or $F_2(\xi) = 0$.
        To determine which one, one finds the sign of
        \begin{equation}
            f_0 = - \sum_{i=1}^{M-1} \sqrt{1-\frac{\lambda_i^2}{\lambda_M^2}} + M - 2.
        \end{equation}
        If $f_0 \geq 0$, solve $F_1(\xi) = 0$ and if $f_0 < 0$, solve $F_2(\xi) = 0$.
        \begin{itemize}
            \item If one has to solve $F_1(\xi) = 0$, $F_1$ has exactly one root $\xi_0$ and $g_\text{max} = g_1(\xi_0)$.
            \item If one has to solve $F_2(\xi) = 0$, $F_2$ has exactly two roots: one at $\xi=0$ and one at $\xi_0 \neq 0$. 
            The maximal separability eigenvalue is given by $g_\text{max} = g_2(\xi_0)$.
        \end{itemize}
        In both cases, the search interval is restricted to $0 < \xi \leq \frac{1}{\lambda_M^2} \equiv \xi_{\mathrm{max}}$. 

        \begin{figure}[t]
            \centering
            \includegraphics[width=\linewidth]{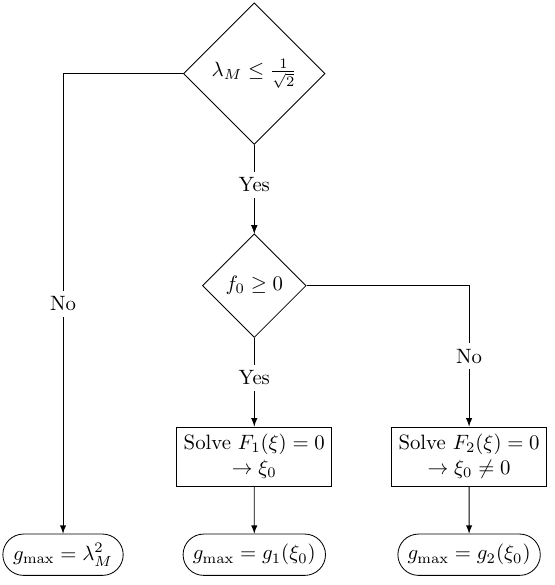}
            \caption{%
                Workflow for solving the separability eigenvalue equations for generalized W states \cref{eq:WState}, described by equation \cref{eq:SEVGeneral}.
                For details, see the main text.
            }\label{fig:OnePhotonFlowChart}
        \end{figure}

        This procedure directly allows for an algorithmic interpretation, which is represented as a flowchart in \cref{fig:OnePhotonFlowChart}.
        Solving the equation $F_1(\xi) = 0$ can be done with a standard root-finding algorithm, like the Newton method and a secant method, as there is only one root on a compact search interval $[0, \xi_\text{max}]$.
        When solving $F_2(\xi) = 0$, there is the trivial root at $\xi=0$ and a nonzero root $\xi_0$.
        To isolate the latter, we reduce the search interval.
        By Rolle's theorem, $\partial_\xi F_2(\xi)$ has at least one root $\tilde\xi$ with $0 < \tilde\xi<\xi_0$.
        Thus, we first solve $\partial_\xi F_2(\xi) = 0$ to find $\tilde\xi$, which is used to find the only root $\xi_0$ of $F_2$ in $[\tilde\xi,\xi_\text{max}]$.
        
        In summary, a maximum of two roots can be determined to find $g_\text{max}$ of any W state.
        The number of modes $M$ enters only through the number of terms in $F_{1/2}$ and $g_{1/2}$.
        Thus, the whole algorithm scales approximately like $O(M)$, making it computationally efficient.
        
        With the above technique, we straightforwardly recover the known result for the symmetric W state ($\lambda_i = 1/\sqrt{M}$)
        \begin{equation}\label{eq:gmaxMax}
            E_{g,\text{max}} = 1- \left( \frac{M-1}{M} \right)^{M-1}.
        \end{equation}
        For example, see also Eq. (15) in \cite{weiGeometricMeasureEntanglement2003} (where $n = M$ and $k = 1$).
        For a fixed $M$, this is the largest $E_g$ for all possible W states, hence $E_{g,\text{max}}$.
        A detailed proof is given in \cref{app:EgMaxForWState}.

        \subsubsection{General partitions}\label{sec:PartitionsOnePhoton}

            We now show how the method introduced above can be applied to find the entanglement for any partition.
            Consider a general W state
            \begin{equation}
                |\psi\rangle = \sum_{i=1}^M \alpha_i |1_i\rangle,
            \end{equation}
            which lives in the Hilbert space $\mathcal{H}_1\otimes\cdots\otimes \mathcal{H}_M$ with $\mathcal{H}_i = \spann \{ |0\rangle, |1\rangle \}$ for at most one photon. 
            We now introduce the general partition $\mathcal{I}_1:\cdots:\mathcal{I}_K$ of $\{1,\ldots,M\}$, where $K < M$.
            The Hilbert space with respect to the general partition is $\mathcal{H}'_1 \otimes \cdots \otimes \mathcal{H}'_K$ with $\mathcal{H}'_k = \spann \big( \{|0\ldots0\rangle\} \cup \{|1_i\rangle \,|\, i=1,\ldots,|\mathcal{I}_k|\} \big)$, where $k=1,\ldots,K$.
            This motivates the following definition
            \begin{equation}
                |1'_k\rangle = \frac{1}{\sqrt{\sum_{i\in\mathcal{I}_k} |\alpha_i|^2}} \sum_{i \in \mathcal{I}_k} \alpha_i |1_i\rangle,
            \end{equation}
            which can be used to rewrite $|\psi\rangle$ as a W state with $K$ modes:
            \begin{equation}\label{eq:onePhotGeneralPartition}
                |\psi\rangle = \sum_{k=1}^K \left(\sqrt{\sum\nolimits_{i\in\mathcal{I}_k} |\alpha_i|^2}\right) |1'_k\rangle.
            \end{equation}
            As this state is again a W state, we can utilize the method form the previous section to compute $g_\text{max}$ and hence $E_g$.

\section{Applications}\label{sec:Applications}

    \subsection{Random networks}\label{sec:Random}

        Before focusing our attention to the entanglement dynamics of specific QWs, it is worthwhile to consider more general properties of entanglement in LONs.
        Specifically, we consider the output of an LON with a randomly selected unitary. 
        We achieve this by choosing a Haar-random unitary $U$ and  computing the entanglement of the state $\hat{\varphi}(U)|\psi_0\rangle$ (see \cref{sec:LONs}).
        Selecting a random unitary matrix is explained, for example, in Refs.\ \cite{mezzadriHowGenerateRandom2007,ozolsHowGenerateRandom2009,mullerNoteMethodGenerating1959}.
        Thereby, we are going to observe the emergence of entanglement typicality \cite{dahlstenEntanglementTypicality2014}.

        \begin{figure}[t]
            \centering
            \includegraphics[width=\linewidthhalf]{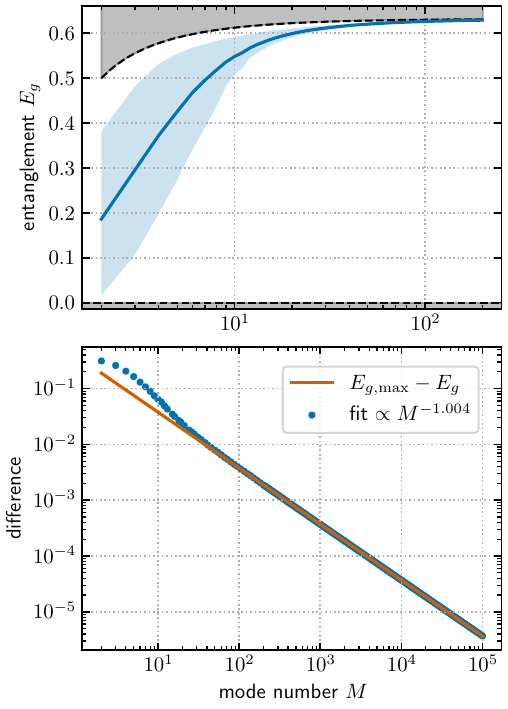}
            \caption{%
                The top panel shows the mean entanglement and the central $68\%$ interval for $5\,000$ uniformly, randomly generated W-states for various mode numbers.
                The dashed black line marks the maximum attainable entanglement $E_{g,\mathrm{max}}$ for a given mode number, according to \cref{eq:gmaxMax}.
                The bottom panel shows the convergence of the mean value to the minimum value in a log-log plot.
                The line represents a linear fit for $M\geq 50$ in the log-log plot, which corresponds to a power law fit for the data.
                The coefficient of determination is $R^2 = 0.999\,987$.
                Mind the different scales of the horizontal axes in both plots.
            }\label{fig:WStateRandom}
        \end{figure}

        We examine the full separability of a single-photon state using the method presented in \cref{sec:FullSepOnePhoton}.
        Choosing a random LON is equivalent to choosing a random W-state, which, in turn, is equivalent to choosing a random point on the $(M-1)$-sphere for $M$ modes.
        This means, instead of generating a full random unitary, it suffices to generate random normalized vectors in $\mathbb{R}^M$ such that their distribution is rotationally invariant \cite{mullerNoteMethodGenerating1959}.
        For different numbers of modes ranging from $M=2$ to $M=10^5$, we generate $5\,000$ random networks each and compute the mean entanglement $\overline{E_g}$ and the central $68\%$ interval, rather than the standard deviation, to emphasize the asymmetry of the distribution.
        The result can be seen in the top of \cref{fig:WStateRandom}.
    
        We observe a convergence towards the maximum $E_g$ for large $M$, i.e., the maximally entangled single-photon state in \cref{eq:gmaxMax}.
        This can be explained geometrically.
        Sampling uniformly on the sphere $S^{M-1}$ gives an expectation value of $1/M$ for the square of each coefficient $\lambda_i$ (i.e., coordinate) with a variance of $1/M$ \cite{mardiaDirectionalStatistics2000}. 
        This converges to $\lambda_i^2 = 1/M$ with certainty for $M\to\infty$, from which we conclude that the system exhibits entanglement typicality \cite{dahlstenEntanglementTypicality2014}.
        For an LON in a continuous-variable setting, such as a Gaussian boson sampler \cite{hamiltonGaussianBosonSampling2017}, a similar observation was made in Refs.\ \cite{iosuePageCurvesTypical2023,qiaoEntanglementFullState2023}.
        This convergence of $E_g$ to the maximum $E_{g,\text{max}}$ can also be seen in the bottom panel of \cref{fig:WStateRandom}. 
        The linear fit in the log-log plot shows that the convergence follows a power law;
        the fit seems to approach $(\rme M)^{-1}$.

    \subsection{Bipartite entanglement dynamics for a QW on the infinite line} \label{sec:CoinEntLine}

        In this section, we derive an exact expression for the entanglement dynamics of a QW on the line as another main result. 
        The bipartition under consideration is given by separating the states where the coin is in the `$-$' state from those where the coin is in the `$+$' state.
        Furthermore, we relate the entanglement for this bipartition to the von Neumann entropy.

        We select initial conditions for a localized state (at zero) and an arbitrary coin state,
        \begin{equation}\label{eq:LocalizedICs}
            |\psi_0\rangle = \left( \cos\frac{\theta}{2} |-\rangle + \rme^{\rmi \phi} \sin\frac{\theta}{2} |+\rangle  \right) \otimes |0\rangle .
        \end{equation}
        We take $\mathcal P=\mathcal I_+:\mathcal I_-$, where $\mathcal{I}_\pm = \{(\pm,x)\}_{x \in \mathbb{Z}}$.
        With respect to this partition, we can use \cref{eq:onePhotGeneralPartition} to find the reduced state.
        Further, it is convenient to work in the Fourier basis \cite{portugalQuantumWalksSearch2018}:
        \begin{equation}
            |\Psi(n)\rangle = \sum_{j = 0,1} \int_{-\pi}^{\pi} \frac{\mathrm{d}k}{2\pi} \tilde\psi_j(k,n) |j\rangle|\tilde k\rangle, 
        \end{equation}
        where $k$ is the wave number, and, for the coin states, we write 
        \begin{equation*}
            |-\rangle \equiv |0\rangle \text{ and } |+\rangle \equiv |1\rangle.
        \end{equation*}
        Note that we use either $|\pm\rangle$ or $|0\rangle$ and $|1\rangle$ for the coin states in this section for convenience.
        Determining the reduced state after $n$ steps,
        \begin{equation}
            |\Phi(n)\rangle = \sqrt{\phi_0(n)} |1',0'\rangle + \sqrt{\phi_1(n)} |0',1'\rangle
        \end{equation}
        amounts to computing the following integrals
        \begin{equation}
            \phi_j(n) = \int_{-\pi}^{\pi} \frac{\mathrm{d}k}{2\pi} |\tilde \psi_j(n,k)|^2 \quad \text{ for } j=0,1 .
        \end{equation}
        Because of normalization, $\phi_0(n) + \phi_1(n) = 1$, it suffices to compute one of the two integrals.
        Without loss of generality, we choose $\phi_1(n)$.

        \paragraph{Solving the integrals.}
        The evolved state can be written as
        \begin{equation}
            |\Psi(n)\rangle = \cos\frac{\theta}{2} |\psi^-(n)\rangle + \rme^{\rmi \phi} \sin\frac{\theta}{2} |\psi^+(n)\rangle,
        \end{equation}
        where $|\psi^\pm(n)\rangle \equiv U^n|\pm,-\rangle$.
        The coefficients for $|\psi^-(n)\rangle$ can be found, e.g., in Eqs.\ (5.41) and (5.42) in Ref.\ \cite{portugalQuantumWalksSearch2018},
        \begin{align}
            \tilde \psi^-_0(n,k) &= \frac{1}{2}\left( 1+ \frac{\cos k}{\sqrt{1+\cos^2k}}  \right) \rme^{-\rmi n \omega_k} \nonumber
            \\
            &\quad + \frac{(-1)^n}{2} \left( 1 - \frac{\cos k}{\sqrt{1+\cos^2k}} \right) \rme^{\rmi n \omega_k},
            \label{eq:QWCoefficientPsi0}
            \\
            \tilde \psi^-_1(n,k) &= \frac{\rme^{\rmi k}}{2\sqrt{1+\cos^2k}}(\rme^{-\rmi n\omega_k}-(-1)^n \rme^{\rmi n \omega_k}),
            \label{eq:QWCoefficientPsi1}
        \end{align}
        where $\omega_k = \arcsin\left(\frac{\sin k}{\sqrt{2}}\right) \in [-\pi/4,\pi/4]$.
        For $|\psi^+(n)\rangle$, we readily arrive at
        \begin{align}
            \tilde \psi^+_0 (n,k) &= (-1)^{n+1} [\tilde \psi^-_1(n,k)]^*
            \\
            \tilde \psi^+_1 (n,k) &= (-1)^n [\tilde \psi^-_0(n,k)]^* .
        \end{align}
        Combining the expressions results in
        \begin{equation}\label{eq:IntegralEqForPhi1}
        \begin{aligned}
            \phi_1(n) &= \frac{1}{2} + \cos\theta\, \underbrace{ \frac{1}{2}\left(2 \int_{-\pi}^{\pi} \frac{\,\mathrm{d}{k}\,}{2\pi} | \tilde \psi^-_1(n,k)|^2-1\right) }_{ I(n) } 
            \\
            &{} + \sin\theta\ \underbrace{ (-1)^n \int_{-\pi}^{\pi} \frac{\,\mathrm{d}{k}\,}{2\pi} \mathrm{Re}( \rme^{-\rmi \phi} \tilde \psi^-_0(n,k) \tilde \psi^-_1(n,k) ) }_{ J(n) }.
        \end{aligned}
        \end{equation}
        We will give a rough sketch of the derivation of $\phi_1(n)$ and state the result below.
        Detailed derivations for both $I(n)$ and $J(n)$ are provided in \cref{app:CoinEntOnLine}.

        \paragraph{Exact solutions.}
        To solve $I(n)$, we rewrite the integrand in terms of Chebyshev polynomials and construct the generating function $G(x) = \sum_{n=0}^\infty I(n) x^n$.
        This results in a standard, rational integral, which is readily solved to give
        \begin{equation}
            G(x) = \frac{1}{2} \left( x\,\frac{1}{1-x} \, \frac{1}{\sqrt{1+x^2}} -1\right).
        \end{equation}
        From this, we read off the result
        \begin{equation}\label{eq:SolutionIntegralIOfn}
            I(n) = \frac{1}{2} \sum_{k=0}^{\left\lfloor \frac{n-1}{2} \right\rfloor} (-1)^k \binom{2k}{k} \frac{1}{4^k} - \frac{1}{2}.
        \end{equation}
        The integral $J(n)$ can be related to $I(n-1)$ and the final result is
        \begin{equation}\label{eq:CoinEntLinePhi1}
            \phi_1(n) = \frac{1}{2} + \cos\theta\,I(n) + \sin\theta \cos\phi\, I(n-1)
        \end{equation}
        and $\phi_0(n) = 1-\phi_1(n)$.

        For the effectively two-dimensional state $|\Phi(n)\rangle$, the entanglement is given by $E_g(n) = \min(\phi_0(n),\phi_1(n))$, as described above.
        For a two-level system, we know $E_g \leq \frac{1}{2}$.
        For the initial parameters $(\theta,\phi)$, we find two regions: one where $\phi_0(n) < \frac{1}{2}$ and one where $\phi_1(n) < \frac{1}{2}$. 
        The boundary is given by the equation
        \begin{equation}\label{eq:CoinEntLineContour}
            \cos\theta\,I(n) + \sin\theta\cos\phi\, I(n-1) = 0.
        \end{equation}

\begin{figure}
            \centering
            \includegraphics[width=7.2cm]{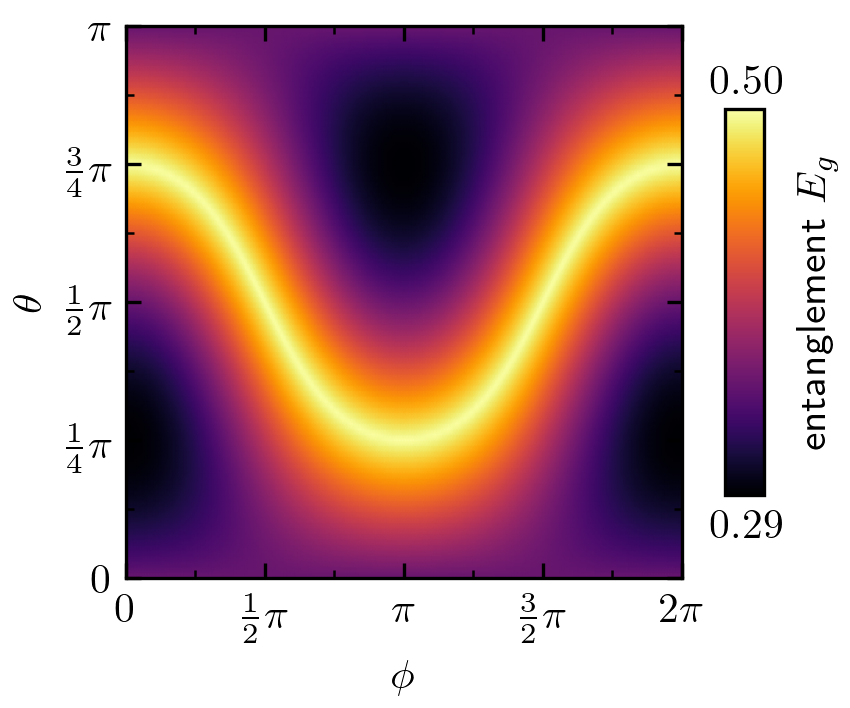}
            \caption{%
                Density plot of the asymptotic entanglement, as given by \cref{eq:AsymptoticEntCoin}, for bipartitions with respect to the coin degrees of freedom.
                The contour follows the form of the great circle, \cref{eq:GreatCircle}, when projected onto the Bloch sphere parametrized by $\theta$ (latitude) and $\phi$ (longitude).
            }\label{fig:CoinEntanglementAsymptoticDensity}
        \end{figure}

        \paragraph{Asymptotic analysis.}
        With the previously derived form, it is straightforward to study the asymptotic case for a QW on the line.
        Consider the power series expansion
        \begin{equation}
            f(x) = \frac{1}{\sqrt{1+x}} = \sum_{k=0}^{\infty} (-1)^k \binom{2k}{k} \frac{1}{4^k} x^k.
        \end{equation}
        From this, it becomes clear that $\lim_{n\to\infty} I(n) = \frac{f(1)-1}{2} = \frac{\sqrt{2}-1}{2\sqrt2}$ and subsequently
        \begin{equation}\label{eq:AsymptoticEntCoin}
            \lim_{n\to\infty} \phi_1(n) = \frac{1}{2} - \frac{\sqrt{2}-1}{2\sqrt2}(\cos\theta+\sin\theta\cos\phi).
        \end{equation}
        A density plot of this function is given in \cref{fig:CoinEntanglementAsymptoticDensity}.
    
        We can find the maximum entanglement with respect to the coin degrees of freedom in the asymptotic limit with \cref{eq:CoinEntLineContour}.
        The maximum entanglement is obtained by starting in the initial conditions that satisfy
        \begin{equation}\label{eq:GreatCircle}
            \cos\theta + \sin\theta\cos\phi = 0.
        \end{equation}
        We recognize this as a great circle \cite{weissteinGreatCircle} on the Bloch sphere, passing through the two points given by $(\theta = \pi/2, \phi=\pi/2)$ and $(\theta=3\pi/4,\phi=0)$, which correspond to the two states
        \begin{equation}
            \frac{|0\rangle+\rmi |1\rangle}{\sqrt{2}} \ \text{ and } \ \frac{\sqrt{2-\sqrt{2}}}{2} |0\rangle + \frac{\sqrt{2+\sqrt{2}}}{2} | 1\rangle,
        \end{equation}
        respectively.
        Interestingly, because $I(2n-1) = I(2n)$, the same contour gives the maximum entanglement for all even time steps, $E_g(2n)=\frac{1}{2}$.
        
        \paragraph{Discussion.}
        In this section, we studied the (bipartite) entanglement between the coin degrees of freedom. 
        As mentioned earlier, the quantity that is usually studied in the context of entanglement in quantum walks is the von Neumann entropy $S_E$ for the partition $\mathcal{P}_\text{coin} : \mathcal{P}_\text{position}$, cf.\ Refs.\ \cite{carneiroEntanglementCoinedQuantum2005,abalQuantumWalkLine2006,ideEntanglementDiscretetimeQuantum2011,ortheyAsymptoticEntanglementQuantum2017}.
        In this case, one usually only considers the asymptotic limit $n \to \infty$.
        The integrals we derived in \cref{eq:IntegralEqForPhi1} can, however, also be used to explicitly compute $S_E$ analytically for all time steps;
        see \cref{app:CoinEntOnLine}.
        We find a strong dependence of the entanglement $E_g$ on the initial coin state.
        However, the asymptotic value of $S_E$ is independent of the initial coin state \cite{carneiroEntanglementCoinedQuantum2005,abalQuantumWalkLine2006}.
        This highlights the importance of the chosen partition and entanglement measure, and why the von Neumann entropy with the usual partition $\mathcal{P}_\text{coin} : \mathcal{P}_\text{position}$ does not capture the full entanglement dynamics.
        Furthermore, the von Neumann entropy cannot be used directly to measure the entanglement of a noisy (i.e., realistic) state.
        The geometric entanglement $E_g$ computed in this section, however, can be used to construct entanglement witnesses, which also detect entanglement under the influence of noise.
    
    \subsection{Multipartite entanglement dynamics for a QW on a circle}\label{sec:EntanglementCircle}

        Next, we consider the entanglement dynamics of a QW with a single walker and periodic boundary conditions.
        We study the general evolution, the impact of initial conditions, and various partitions.
        
        \subsubsection{Simple initial conditions}

\begin{figure}
                \centering
                \includegraphics[width=\linewidthhalf]{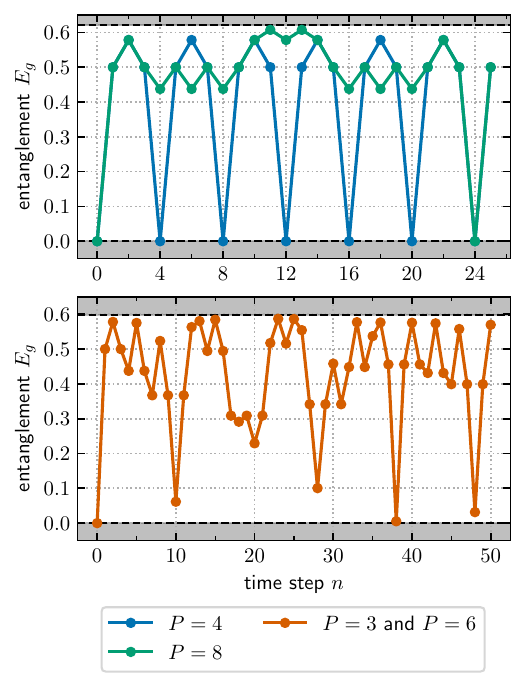}
                \caption{%
                    Entanglement dynamics $E_g(n)$ for different number of positions $P$.
                    The gray areas mark regions of $E_g$ that cannot be attained;
                    see \cref{eq:gmaxMax}.
                    Note that the entanglement dynamics for $P=3$ and $P=6$ are identical.
                }\label{fig:HadWalkSimpleICDifferentP}
            \end{figure}

            We start by considering the full separability $\mathcal{S} = \{1\}:\cdots:\{M\}$ of the separable initial state $|\psi_0\rangle =|-,0\rangle$.
            For $M=2P$ modes, the entanglement evolution for different values of $P$ is shown in \cref{fig:HadWalkSimpleICDifferentP}.
            
            \begin{figure*}
                \centering
                \includegraphics[width=\linewidthfull]{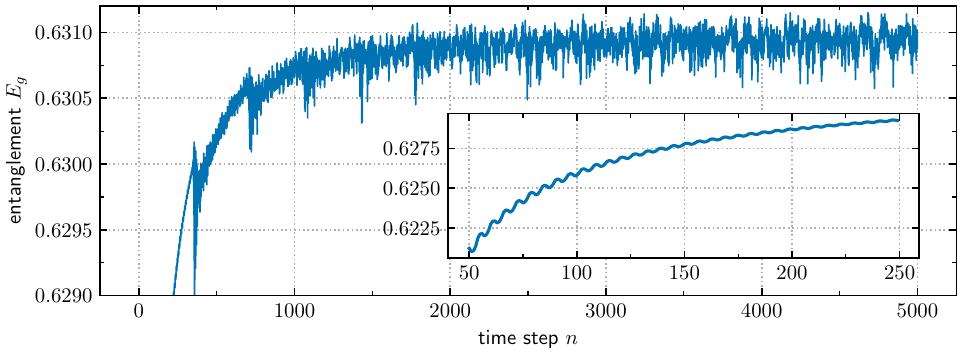}
                \caption{%
                    Entanglement dynamics $E_g(n)$ for a QW of size $P=500$.
                    The inset highlights the regular dynamics before the boundaries are reached at time step $n=250$.
                    Local minima are always spaced apart by $6$ steps, except for three cases. The irregular behavior for $n > 250$ is limited to a rather small region of $E_g$.
                }\label{fig:HadWalkSimpleIC:P500}
            \end{figure*}
            
            Firstly, we note that the dynamics for $P=1$, $P=2$, $P=4$, and $P=8$ is periodic ($P=1,2$ not shown) with the periods $2$, $2$, $4$, and $24$, respectively.
            This is consistent with the periods for the Hadamard walk of $2,2,8,$ and $24$;
            see, e.g., Table 1 in Ref.\ \cite{tregennaControllingDiscreteQuantum2003}.
            For $P=4$, we have $\hat T^4|-,0\rangle=|-,2\rangle$, which explains its period of $4$ instead of $8$.
            As proven in Ref.\ \cite{konnoPeriodicityHadamardWalk2017}, the Hadamard walks with $P\in\{1,2,4,8\}$ are the only periodic cases.
            By contrast, the non-periodic walks given in \cref{fig:HadWalkSimpleICDifferentP,fig:HadWalkSimpleIC:P500} exhibit quasi-periodic behavior.
            This has previously been noted and studied in Ref.\ \cite{rohdeEntanglementDynamicsQuasiperiodicity2012} for a Hadamard walk with reflective boundary conditions.

            The evolution for $P=3$ and $P=6$ in \cref{fig:HadWalkSimpleICDifferentP} are identical.
            This reflects a general property of the Hadamard walk with periodic boundary conditions.
            For odd $P$, the entanglement dynamics for $P$ and $2P$ are the same because of parity conservation and breaking.
            On the line, the QW conserves parity. 
            If the walker initially occupies only even (likewise, only odd) positions $x$, it always occupies positions of that parity, alternating at each time step.
            When introducing periodic boundary conditions, this parity conservation is conserved for even $P$, yet broken for odd $P$.
            Let $|\psi\rangle$ denote the state of the QW for odd $P$ at some arbitrary time step.
            After reordering the modes, the quantum state of the QW for $2P$ positions can be written as $|0\rangle^{\otimes P}|\psi\rangle$.
            Since vacuum modes do not contribute to the geometric measure of entanglement, the entanglement evolution for odd $P$ and $2P$ are identical.
            Although this property is specific to periodic boundary conditions, it indicates that simply increasing the system size does not necessarily increase the dynamics' complexity.

            To highlight the potential of our method, we computed the full entanglement dynamics of a very large system with $P=500$ for $5\,000$ time steps, as shown in \cref{fig:HadWalkSimpleIC:P500}.
            Several features are worth noting.
            While the walk still exhibits the quasi-periodic `chaotic' nature, similar to smaller $P$, it seems to converge to a small region of relatively high entanglement.
            For $P=3,6$ (top panel of \cref{fig:HadWalkSimpleICDifferentP}), the walk regularly approached $E_g \approx 0$, before revisiting a much higher value, again.
            In Ref.\ \cite{rohdeEntanglementDynamicsQuasiperiodicity2012}, it was noted that complex quasi-periodic dynamics \cite{ideEntanglementDiscretetimeQuantum2011} can emerge for systems of finite size.
            It was concluded that `simply waiting for a very long time' does not necessarily guarantee a high amount of entanglement. 
            We agree with this statement only for smaller systems as large systems seem to restrict their rapid oscillations to a relatively small region of high entanglement.
            Furthermore, as also noted in Ref.\ \cite{rohdeEntanglementDynamicsQuasiperiodicity2012}, the irregular structure only emerges after the walk hits the boundary.
            Because of the larger system size of $P=500$, instead of $P=100$ in Ref.\ \cite{rohdeEntanglementDynamicsQuasiperiodicity2012}, we can observe the regular behavior for small times more clearly.
            While the left-propagating and right-propagating parts of the walk begin to interfere at $n= P/2 = 250$, the breakdown of the regular dynamic does not happen immediately.
            This feature is further discussed in \cref{sec:arbParts}.
            Additionally, we note the peculiar regular oscillation present before the boundary conditions are applied.
            The peaks of this oscillation have an almost perfect period of $6$, with only three occurrences where the QW slightly deviates from this behavior.
            To the best of our knowledge, this phenomenon has not been reported previously, and its underlying mechanism remains unknown.

        \subsubsection{Arbitrary localized initial conditions}

            There are numerous ways to change the parameters of the walk, which include more walkers, a different topology, different initial conditions, a different coin dimension, and more.
            We now focus on the influence of the initial conditions on the walk as we did in \cref{sec:CoinEntLine}.
            That is, we choose the localized initial condition given by \cref{eq:LocalizedICs} and numerically study the multipartite entanglement of the QW on a circle.

            The entanglement dynamics are independent on the initial condition as long as the initial state is of the form $|c,x\rangle$, with $c=\pm$ and $x=0,\ldots,P-1$. 
            The independence of the starting position is clear because of the rotational (in the case of $P\to\infty$ translational) symmetry of the topology under study.
            For simplicity, we only consider the Hadamard coin, which has a balanced coin.
            Flipping the initial coin state $-\to+$ simply mirrors the walk \cite{kempeQuantumRandomWalks2003}.
            Thus, the interesting initial conditions are superpositions of states of the form $|c,x\rangle$ for $c\neq\pm$.

\begin{figure}
                \centering
                \includegraphics[width=7.8cm]{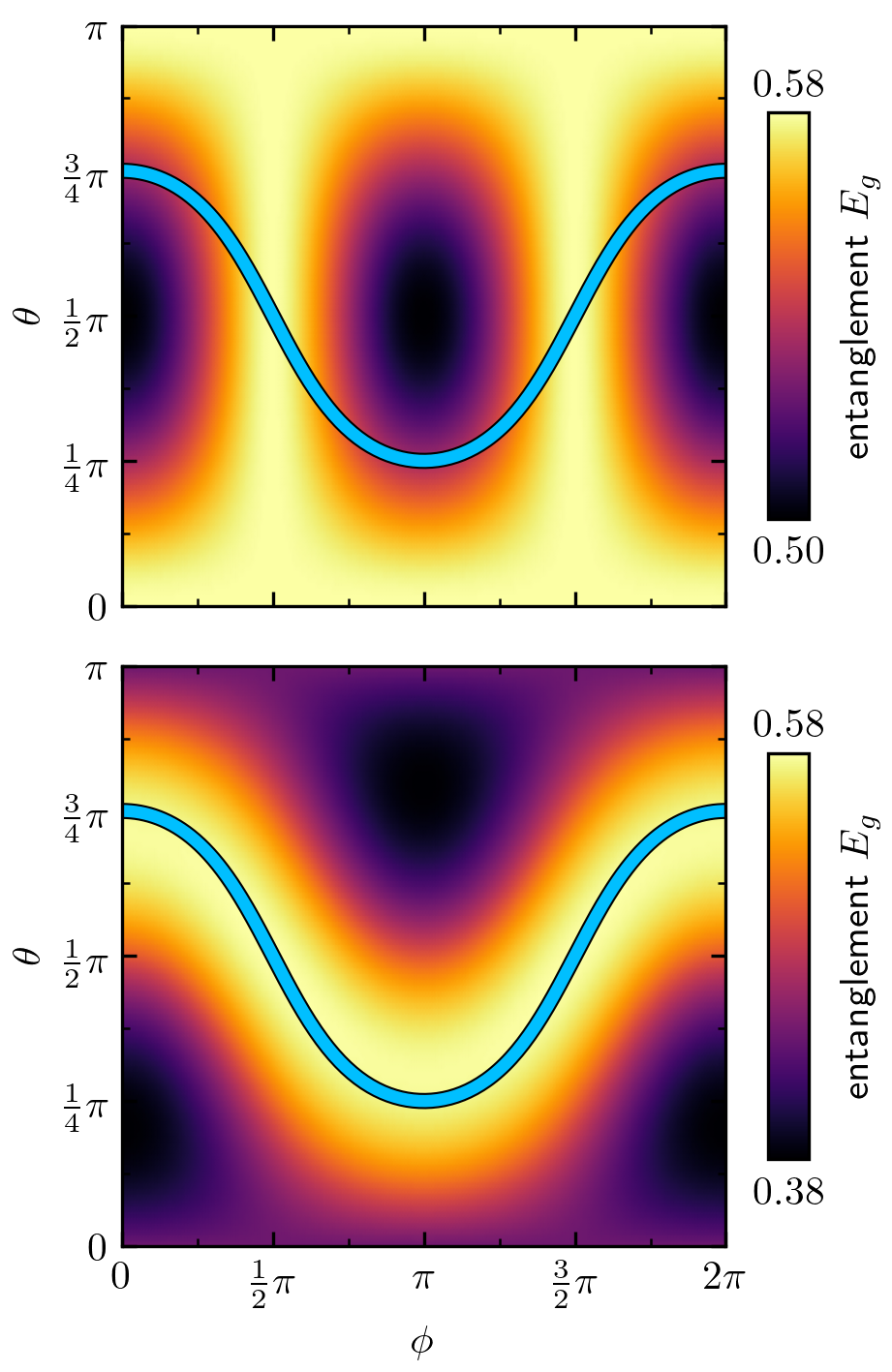}
                \caption{%
                    Entanglement for an arbitrary localized initial state \cref{eq:LocalizedICs} for a QW with $P=5$.
                    The top panel shows the entanglement at time $n=2$, and the bottom panel shows the entanglement at time $n=4$.
                    To highlight the structures in the two panels, different colorbar ranges are used.
                    The blue, thick line highlights the great circle given by \cref{eq:GreatCircle}. 
                    Note the similarity of the bottom panel to \cref{fig:CoinEntanglementAsymptoticDensity}.
                }\label{fig:LocalizedICsDensityPlots}
            \end{figure}

            As mentioned above, we have localized initial conditions with an arbitrary initial coin state of the form given in \cref{eq:LocalizedICs}.
            They are especially interesting from an experimental point of view as they are simply achievable via wave plates in photonic implementations that use the polarization as the coin \cite{broomeDiscreteSinglePhotonQuantum2010}. 
            For a small network, $P=5$, the entanglement for two different time steps as a function of $\theta$ and $\phi$ is shown in \cref{fig:LocalizedICsDensityPlots}.

            We have also shown the importance of the great circle on the Bloch sphere that was given in \cref{eq:GreatCircle}.
            We observe that this line gives a good approximation for the local maximum for $n=4$, but not for $n=2$, cf.\ \cref{fig:LocalizedICsDensityPlots}.
            As it turns out, when considering larger networks, \cref{eq:GreatCircle} is a good approximation for the local maximum for most time steps.
            The larger the network, the better this approximation gets. 
            This leads us to conjecture that \cref{eq:GreatCircle} describes the local maximum for the full entanglement for a QW on the line in the asymptotic case $n\to\infty$.
            Also note the similarity of the bottom panel in \cref{fig:LocalizedICsDensityPlots} to the findings reported in Figure 4 of Ref.\ \cite{carneiroEntanglementCoinedQuantum2005}.
            
            The choice $\theta=\pi/2,\, \phi=\pi/2$ for the angles that define the initial coin corresponds to a symmetric walk \cite{kempeQuantumRandomWalks2003} as the evolutions of the two components of the initial state, $|-,0\rangle$ and $\rmi|+,0\rangle$, do not interfere. 
            Hence, we expect this walk to always yield the maximum entanglement.
            This is indeed the observed behavior for all graph sizes and time steps we have tested.

            \begin{figure*}
                \centering
                \includegraphics[width=\linewidthfull]{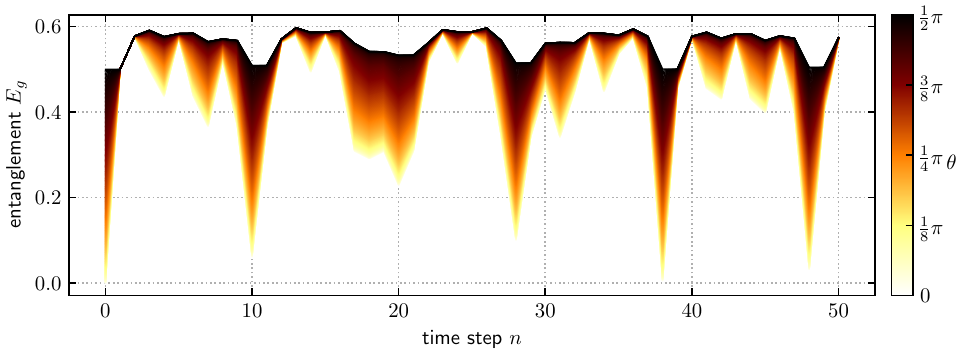}
                \caption{Entanglement evolution for a QW with $P=3$ for fixed $\phi=\pi/2$ and varying $\theta \in [0,\pi/2]$ in the initial state \cref{eq:LocalizedICs}. Following a fixed color gives the entanglement evolution for the state of that particular initial state.}
                \label{fig:LocalizedICsVaryingTheta}
            \end{figure*}

            We can visualize the increase in entanglement for the non-mixing case nicely, when we smoothly interpolate between $|\psi_0\rangle=|-,0\rangle$ to $|\psi_0\rangle = (|-,0\rangle + \rmi |+,0\rangle)/\sqrt{2}$ by setting $\phi=\pi/2$ and varying $\theta$ in \cref{eq:LocalizedICs}.
            The result can be seen in \cref{fig:LocalizedICsVaryingTheta}.
            There, we observe a monotonous increase in $E_g$ at every time step for increasing $\theta = 0 \to \pi/2$.
            We can restrict ourselves to $\theta \leq \pi/2$ as the interval $[\pi/2,\pi]$ is redundant.
            Only the absolute values of the coefficients of the W state are relevant to the entanglement and the Hadamard walk has a balanced coin.
            The network size of the walk was chosen small on purpose as the effect is most pronounced here.

        \subsubsection{Different partitions}\label{sec:arbParts}

            \begin{figure*}
                \centering
                \includegraphics[width=\linewidthfull]{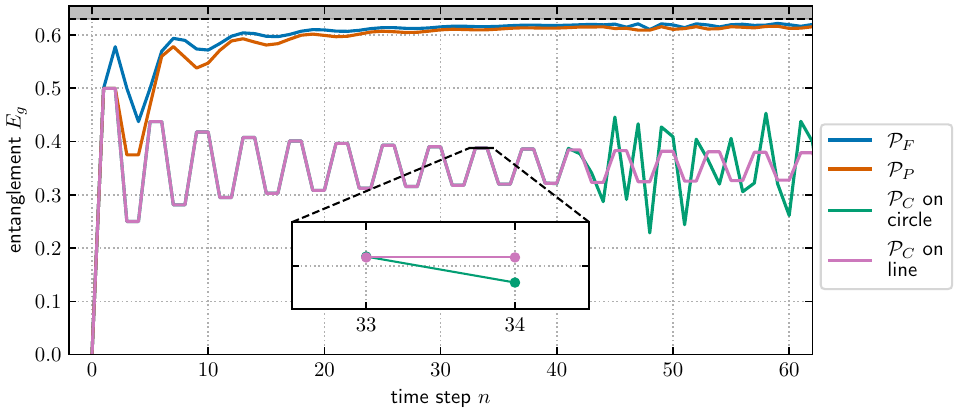}
                \caption{%
                    Entanglement of the Hadamard walk of size $P=64$ for different partitions.
                    The entanglement between all modes is shown in blue, the entanglement between the two coin degrees of freedom in green, and the entanglement between the positions in orange.
                    The entanglement between the coin degrees of freedom for a QW walk on the line is shown in purple.
                    At $n=P/2=32$, the periodic boundary conditions have an impact for the first time, as the left- and right-going parts of the walk begin to interfere.
                }\label{fig:HadWalkDifferentPartitions}
            \end{figure*}

            We now consider different partitions for the QW, as noted in \cref{sec:PartitionsOnePhoton}.
            In this section, we consider three partitions:
            \begin{itemize}
                \item full separation $\mathcal{P}_F=\mathcal{I}_{\pm,0}:\cdots: \mathcal{I}_{\pm,P}$, with $\mathcal{I}_{c,x} = \{(c,x)\}$,
                \item separation w.r.t. the position $\mathcal{P}_P = \mathcal{I}_0 : \cdots: \mathcal{I}_P$, with $\mathcal{I}_x = \{(-,x),(+,x)\}$,
                \item separation w.r.t. the coin $\mathcal{P}_C = \mathcal{I}_- : \mathcal{I}_+$, with $\mathcal{I}_\pm = \{(\pm,x)\}_{x=0}^P$.
            \end{itemize}
            For the separation w.r.t. the coin $\mathcal{I}_- : \mathcal{I}_+$, we derived analytical solutions for $P\to\infty$ in \cref{sec:CoinEntLine}.
            Here, we return to the QW on a circle $P < \infty$.
            The entanglement dynamics for the walk are shown in \cref{fig:HadWalkDifferentPartitions}.
            Firstly, we can clearly see the effect of the periodic boundary conditions for $\mathcal{P}_C$.
            At $n=P/2=32$, the left-propagating part of the QW and the right-propagating part interfere for the first time.
            Similarly to the behavior in \cref{fig:HadWalkSimpleIC:P500}, the change from a regular dynamic to an irregular (i.e., quasi-periodic) dynamic does not happen abruptly but in smaller increments.
            Using the result in \cref{eq:CoinEntLinePhi1}, we can directly compare the entanglement for $\mathcal{P}_C$ on the circle with that on the line.            
            Numerically, we know that the two lines already deviate at $n=33$ with a difference of about $2\times10^{-10}$. 
            For $n=34$, the difference becomes apparent in the zoomed-in section in \cref{fig:HadWalkDifferentPartitions}, with a difference of about $7 \times 10^{-9}$, further increasing with every step.
            At around $n=42$, the deviation becomes visible by eye in the full plot, transitioning from a fairly regular to a highly irregular entanglement pattern.

\section{Multi-walker scenarios}\label{sec:MultiplePhotons}

    So far, we have focused on the single-photon (single-walker) case, $N=1$.
    In this section, we consider $N > 1$.
    The complexity significantly increases, and the method that leads to the efficient algorithm in \cref{sec:EntanglementWState} cannot be extended straightforwardly to the case of multiple photons.
    Here, we present two different aspects of entanglement.
    We firstly comment on our expectation for the entanglement typicality for the geometric measure of entanglement.
    Then, we introduce a method to compute the bipartite entanglement of arbitrary networks with $N, M \in \mathbb{N}$.
    We discuss its application in studying random networks via the genuine multipartite entanglement.

    \subsection{Geometric measure of entanglement}

        In \cref{sec:Random}, we found the convergence of $E_g$ to the maximum entanglement $E_{g,\text{max}}$ in the case of a single photon, as given by \cref{eq:gmaxMax}. 
        We noted that this maximum is achieved by the equally weighted W state.
        We will extend this reasoning to propose the limit of $E_g$ for $N > 1$ as $M \to \infty$.
        
        For large networks $M \gg N$, all contributions of states with two or more photons in the same mode are suppressed \cite{aaronsonComputationalComplexityLinear2011}.
        Hence, the state can be well approximated by a Dicke state\footnote{That is, states of the form $\sum_{\text{permutations}} |0_1\ldots0_k 1_{k+1} \ldots1_M\rangle$, where $k$ is constant \cite{weiGeometricMeasureEntanglement2003}.}. 
        For a single photon, we saw that a random state for large networks is close (entanglement-wise) to the equally weighted W state.
        Due to the Haar-random choice of the network's unitary, we expect an input of the form $|\psi_0\rangle = |1_1\ldots1_N0_{N+1}\ldots0_M\rangle$ to be mapped to a state that is close to a symmetric Dicke state.
        The geometric measure of entanglement for a symmetric Dicke state is given in Eq.\ (15) in Ref.\ \cite{weiGeometricMeasureEntanglement2003} as
        \begin{equation}\label{eq:EgMaxDicke}
            E_{g,\text{max}} = 1 - \binom{M}{N} \left( \frac{N}{M} \right)^N \left( \frac{M-N}{M} \right)^{M-N},
        \end{equation}
        and we thus conjecture
        $
            E_g\left( \hat\varphi(U) |1\ldots10\ldots0\rangle \right) \approx E_{g,\text{max}}
        $
        for large $M$ and Haar-random $U \in \mathrm{U}(M))$, analogously to the single-photon scenario.

    \subsection{Bipartitions and genuine multipartite entanglement} \label{sec:Bipartite}

        Next, we consider the entanglement for an arbitrary number of photons $N$ and an arbitrary mode number $M$, yet with the restriction to bipartitions.
        That is, we consider partitions $\mathcal{P} = \mathcal{I}_1:\mathcal{I}_2$ of $\{1,\ldots,M\}$.
        
        The entanglement of a state is invariant under mode reordering; 
        hence, w.l.o.g., we can assume $\mathcal{I}_1 = \{1,\ldots,k\}$ and $\mathcal{I}_2 = \{k+1,\ldots,M\}$ for some $1\leq k < M$.
        To compute the entanglement of a bipartition, we utilize the Schmidt decomposition \cite{nielsenQuantumComputationQuantum2012}
        \begin{equation}\label{eq:SchmidtGeneral}
            |\psi\rangle = \sum_{i=1}^r \sigma_i |\alpha_i \rangle |\beta_i\rangle, 
        \end{equation}
        with the Schmidt rank $r$, the Schmidt coefficients $\sigma_i$, and the Schmidt bases $\{|\alpha_i\rangle\}$ and $\{|\beta_i\rangle\}$.
        The largest Schmidt coefficient is equal to the sought separability eigenvalue, i.e., $g_\text{max} = \max_i \sigma_i^2$ \cite{sperlingNecessarySufficientConditions2009}.

        We start with a general output state of an LON, as given by \cref{eq:OutputLON},
        \begin{equation}
            |\psi\rangle = \sum_{\vec n:\, |\vec n| = N} \psi_{\vec n} |n_1\ldots n_k\rangle |n_{k+1}\ldots n_M\rangle.
        \end{equation}
        We can split the sum into orthogonal subspaces, where $t$ photons are in the subspace $\mathcal{I}_1$ and $N-t$ photons are in the subspace $\mathcal{I}_2$:
        \begin{equation}\label{eq:SchmidtDecompSplitState}
            |\psi\rangle = \sum_{t=0}^N \sum_{\substack{ n_1,\ldots,n_k: \\ |\vec n| = t }} \sum_{\substack{ m_1,\ldots,m_{M-k}: \\ |\vec m| = N-t }} \psi_{\vec n, \vec m} |\vec n\rangle |\vec m\rangle.
        \end{equation}
        To find the Schmidt decomposition, it suffices to compute the Schmidt decomposition of the individual subspaces
        \begin{equation}
            \sum_{\substack{ n_1,\ldots,n_k: \\ |\vec n| = t }} \sum_{\substack{ m_1,\ldots,m_k: \\ |\vec m| = N-t }} \psi_{\vec n, \vec m} |\vec n\rangle |\vec m\rangle = \sum_{i=1}^{r^{(t)}} \sigma_i^{(t)} |\alpha_i^{(t)}\rangle |\beta_i^{(t)}\rangle,
        \end{equation}
        which immediately yields
        \begin{equation}
            |\psi\rangle = \sum_{i=1}^r \sigma_i |\alpha_i \rangle |\beta_i\rangle = \sum_{t=0}^n \sum_{i=1}^{r^{(t)}} \sigma_i^{(t)} |\alpha_i^{(t)}\rangle |\beta_i^{(t)}\rangle.
        \end{equation}
        In this form, the maximum separability eigenvalue can be directly extracted, being
        \begin{equation}
            g_\text{max} = \max_{t=0,\ldots,N} \max_{i=1,\ldots,r^{(t)}} |\sigma_i^{(t)}|^2.
        \end{equation}
        Calculating the singular values of an $\mu\times \nu$ matrix is an $O(\mu\,\nu\min(\mu,\nu))$ operation \cite{golubMatrixComputations2013}.
        For $m$ modes and $n$ photons, the dimensionality of the Hilbert space is $\binom{m-1+n}{n}$.
        For a given pair $k$ and $t$, we thus have $\mu = \binom{k-1+t}{t}$ and $\nu=\binom{(M-k)-1+(N-t)}{N-t}$ and a $\mu\times \nu$ coefficient matrix $\psi_{\vec n, \vec m}$.
        We find that the quantity $\mu\,\nu\min(\mu,\nu)$ is maximized for $k \approx M/2$ and $t \approx N/2$.
        This gives a total worst-case scaling of $O\left( \binom{M/2 - 1 + N/2}{N/2}^3 \right)$.
        Finally, to make statements about the multipartite entanglement using bipartitions, one often employs the genuine multipartite entanglement \cite{tothPracticalMethodsWitnessing2009}.
        Here, the geometric entanglement generalizes to
        \begin{equation}\label{eq:GME}
            G_g \equiv \min_{\mathcal{P} = \mathcal{I}_1 : \mathcal{I}_2} E_g,
        \end{equation}
        where the minimum runs over all possible bipartitions $\mathcal{I}_1 : \mathcal{I}_2$ of $\mathcal{I}$ to account for all separations that define genuine entanglement.

        \subsection{Application to random networks}

        \begin{figure*}
                \centering
                \includegraphics[width=\linewidthfull]{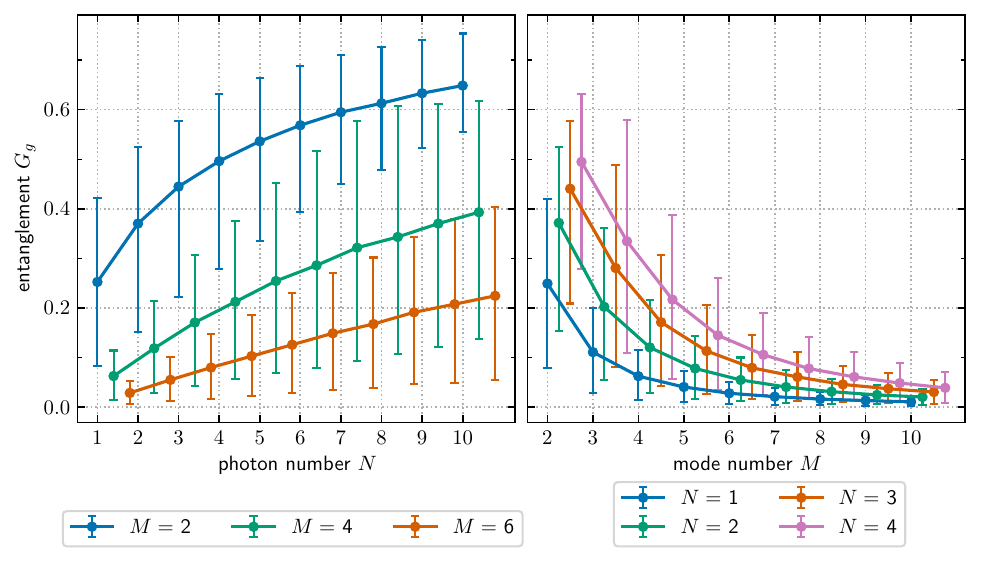}
                \caption{%
                    The genuine multipartite entanglement for $6\,000$ states, evolved according to a uniformly, Haar-randomly generated unitary for various combinations of mode numbers $M$ (left) and number of photons $N$ (likewise, number of walkers; right).
                    The left panel highlights how the mean and central $68\%$ interval (see text) behave when keeping $M$ fixed and increasing $N$.
                    The right panel shows how the mean and central $68\%$ interval behave when keeping $N$ fixed and increasing $M$.
                }\label{fig:BipartiteRandom}
            \end{figure*}
    
            We now apply our previous observations to the genuine multipartite entanglement $G_g$ of Haar-random LONs with more than one photon.
            In \cref{fig:BipartiteRandom}, we consider $G_g$ of states that are propagated from the initial state $|\psi_0\rangle=|N0\ldots0\rangle$ according to a randomly generated $\mathrm{U}(M))$ matrix.
            That is, we compute $G_g(\hat\varphi(U)|\psi_0\rangle)$ for $N$ walkers starting in the $|-,0\rangle$ state.
            We choose this specific initial state, instead of the more commonly seen $|1\ldots10\ldots0\rangle$, because the qualitative behavior of the entanglement does not change.
            The states $|1\ldots10\ldots0\rangle$ are restricted to $M \geq N$, while the states $|N0\ldots0\rangle$ are not. 
            This allows for a larger range of possible combinations of $M$ and $N$, as seen in \cref{fig:BipartiteRandom}.
            
            We plot both the mean and the central $68\%$ interval of $G_g$ in \cref{fig:BipartiteRandom}.
            As discussed in \cref{sec:Random}, the central $68\%$ interval is used to emphasize the asymmetry of the distribution.
            For fixed $M$, we can see an increase in $G_g$ for an increase in $N$, but also an increase in the deviation.
            It is intuitive that a higher number of photons, distributed among the same number of modes, leads to a higher entanglement.
            For fixed $N$, we see that the mean $G_g$ and its deviation both tend to zero for increasing $M$.
            In \cref{sec:Random}, we saw that an increase in the mode number led to an increase in entanglement.
            Here, however, we see the exact opposite.
            This is because the same number of photons spread out over a larger number of modes, which, in general, decreases the modulus of the amplitudes for each mode.
            This makes it much more likely that a mode is close to the vacuum state, i.e., it can `almost' be separated from the other modes.
            We conclude that, for a fixed number of photons, no entanglement typicality emerges with respect to the genuine multipartite entanglement.

            In QWs we often consider large networks with a small number of walkers. 
            As the genuine multipartite entanglement tends to go to zero for these systems, we conclude that it is not a suitable entanglement measure for this situation.
            This is amplified by the fact that for a QW on the line, only the even or odd positions are occupied at any given time, which would result in $G_g = 0$.
            Thus, more advanced measures may be constructed in the future to overcome such limitations.

\section{Conclusion}\label{sec:Conclusion}

    We developed methods to efficiently study the multipartite entanglement dynamics of quantum walks and linear optical networks in various settings.
    For a single photon, we presented an efficient algorithm to compute the geometric measure of entanglement.
    This proved to be a valuable tool for addressing aspects of quantum walks and linear optical networks that previous methods could not fully capture.
    This includes an analytical expression for the entanglement between the coin degrees of freedom of a quantum walk on the line.

    Our approach allowed us to systematically analyze the dynamics of the entanglement.
    We formulated an analytical solution for the complete time evolution for arbitrary localized initial conditions and discussed its asymptotic structure.
    Furthermore, we showed how the solution also leads to exact expressions for the von Neumann entropy between the coin and position degrees of freedom, which is considered often as a measure of entanglement for quantum walks in the literature.

    We numerically investigated single-walker quantum walks on a circle, studying network sizes which are orders of magnitude larger than what was previously possible.
    We observed the previously known quasi-periodic nature of the entanglement in quantum walks of finite size.
    But, because of the efficiency of our algorithm, we were able to study the entanglement at late times for very large system sizes.
    Furthermore, we studied the transition from the regular to the irregular quasi-periodic behavior, showing that the change occurs gradually over multiple time steps, instead of abruptly.
    We also studied the dependence of the entanglement dynamics on the localized initial conditions and found it to be highly sensitive to the initial condition of the coin, which is relevant for experiments with limited control over the initial state.
    This dependency also holds for the asymptotic case, which is in contrast to the von Neumann entropy, where this dependence is not detectable.

    Beside studying specific quantum walks, we considered statistical ensembles of linear optical networks.
    We found the emergence of entanglement typicality for the geometric measure of entanglement in the case of a single photon, and characterized its convergence.
    We additionally considered the case of multiple walkers, relying mostly on bipartitions, to study genuine multipartite entanglement.
    We showed how the constant photon number can be used to reduce the dimensionality of the singular value decomposition.
    As an application, we considered random networks; yet the emergence of entanglement typicality, with respect to the genuine multipartite entanglement, was infeasible because of the increased complexity, limiting all studies of complex multipartite entanglement.

    In this study, we took the Hadamard walk on the line and on a circle as proof-of-concept examples.
    However, the algorithm we presented for a single walker is not restricted to this choice.
    In fact, due to the efficiency of the algorithm, even large networks for a walk on different, higher dimensional topologies can be studied, such as done in Ref.\ \cite{difidioQuantumWalksEntanglement2024}.
    Extending the study performed here to different topologies and completely different walks is certainly one interesting direction for future research.
    Additionally, it would be interesting to know if the method we used in \cref{sec:CoinEntLine}, to study the entanglement between the coin degrees of freedom, can be generalized to, for example, non-local initial conditions and different types of coins.
    Furthermore, non-linear quantum walks are gaining more attention \cite{heldDrivenGaussianQuantum2022} and would make for an interesting extension to which our techniques can be generalized.
    However, the non-constant photon number of such nonlinear processes will make their treatment significantly more challenging.
    
    Another interesting direction would be to revisit the case of multiple walkers.
    The method we used, which is based on bipartitions, has several drawbacks.
    For example, it does not scale favorably with the system size, and, since it is based on bipartitions, it does not capture the full, higher-order multipartite entanglement nature, such as  reported, e.g., in Ref.\ \cite{gerkeMultipartiteEntanglementTwoSeparable2016}.
    Extending the methods for the geometric measure of entanglement to more than a single photon would be desirable.

\section*{Data availability}

    The data that support the findings of this article and the underlying code base used to generate it are openly available \cite{zenodoLink}.

\begin{acknowledgments}
    The authors are grateful to Laura Ares for valuable comments and thank Martin Ammon for fruitful discussions.
    E.K.F.D.\ thanks the Theoretical Quantum Science group for their kind hospitality.
    J.S.\ acknowledges the support through the Deutsche Forschungsgemeinschaft (DFG) via the TRR 142/3 (Project No.\ 231447078, Subproject No.\ C10).
\end{acknowledgments}

\appendix

\section{The uniformly weighted W state and maximized entanglement}\label{app:EgMaxForWState}

    We show that the W state with equal weights,
    \begin{equation}
        |W_M\rangle = \frac{1}{\sqrt{M}} \sum_{k=1}^{M} |1_k\rangle,
    \end{equation}
    is the state with maximum entanglement, for fixed $M$ and $N=1$ walker.
    That is, the maximum separability eigenvalue $g_\text{max}$ is minimized by this state:
    \begin{equation}
        g_\text{max}(|\psi\rangle) \geq g_\text{max}(|W_M\rangle).
    \end{equation}
    We introduce an additional subsystem, giving the following $(M+1)$-dimensional W state
    \begin{equation}
        |\psi(x)\rangle = x |W_M\rangle|1\rangle + \sqrt{1-x^2} |\vec{0}\rangle |1\rangle,
    \end{equation}
    where $x \in [1/\sqrt{2}, 1]$. 
    For $x < 1/\sqrt{2}$, according to \cref{sec:EntanglementWState}, the maximum separability eigenvalue would be $x^2$, which is not the minimum with respect to all possible $x$.
    Next, we show that $g_\text{max}(x) \equiv g_\text{max}(|\psi(x)\rangle)$ is minimized for $x = \frac{M}{M+1}$, which corresponds to the uniformly weighted W state $|W_{M+1}\rangle$.
    We consider the test operator (cf.\ \cref{eq:EntanglementWitnessDefinition})
    \begin{equation}
        \hat L(x) = |\psi(x)\rangle\langle\psi(x)|,
    \end{equation}
    which is of rank one.
    By the cascaded structure theorem \cite{sperlingMultipartiteEntanglementWitnesses2013}, the non-zero solutions of the separability eigenvalue equations of $\hat L$ are identical to the solutions of the operator
    \begin{equation}
        \hat L'(x) = \operatorname{tr}_{M+1} \hat L = x^2 |W_M\rangle\langle W_M | + (1-x^2)|\vec 0\rangle \langle \vec 0|,
    \end{equation}
    which is of rank two.
    In particular, the maximum separability eigenvalue $g_\text{max}$ is identical for $\hat L(x)$ and $\hat L'(x)$.
    In Corollary 5 of Ref.\ \cite{hubenerGeometricMeasureEntanglement2009}, it was shown that for a permutationally invariant positive operator $\hat X$, the maximum
    \begin{equation}
        \max_{|\chi_i\rangle \in \mathcal{H}_i} \langle \chi_1\ldots \chi_M|\hat X|\chi_1\ldots \chi_M\rangle
    \end{equation}
    is attained via a symmetric state, $\forall i:|\chi_i\rangle = |\chi\rangle$.
    Since $\hat L'(x)$ is a permutationally invariant positive operator, the separability eigenvector pertaining to $g_\text{max}$ is symmetric:
    \begin{equation}
        |\vec \chi(y)\rangle = |\chi(y)\rangle^{\otimes M} \equiv \left( y |0\rangle + \sqrt{1+y^2}|1\rangle \right)^{\otimes M}.
    \end{equation}
    We define
    \begin{equation}
        g(x,y) = \langle \vec \chi(y)| \hat L'(x)| \vec \chi(y)\rangle
    \end{equation}
    and want to find $x$ that solves
    \begin{equation}
        \min_x \max_y g(x,y).
    \end{equation}
    Setting $u = x^2$ and $v = y^2$, we find
    \begin{equation}
        g(u,v) = v^M [1-u(M+1)] + M v^{M-1} u.
    \end{equation}
    We maximize with respect to $v$, we find
    \begin{equation}
        \frac{\partial g(u,v)}{\partial v} = 0 \, \Rightarrow \, v = \frac{u(M-1)}{u(M+1)-1}.
    \end{equation}
    We then need to find the minimum of
    \begin{equation}
        g(u) = u \left( \frac{(M-1) u}{u(M+1)-1} \right)^{M-1}.
    \end{equation}
    We get the non-trivial solution
    \begin{equation}
        \frac{\mathrm{d} g(u)}{\mathrm{d} u} = 0 \,\Rightarrow\, u = \frac{M}{M+1}.
    \end{equation}
    It is straightforward to confirm that these $u,v$ do in fact correspond to a minimum and maximum, respectively.
    Hence, we have
    \begin{equation}
        g(x,y) = \langle \vec \chi(y)| \hat L'(x)| \vec \chi(y)\rangle \,\Rightarrow\, x = \sqrt{\frac{M}{M+1}},
    \end{equation}
    which means $|\psi(x)\rangle = |W_{M+1}\rangle$.
    The above induction step from $M$ to $M+1$ completes our proof.

\section{Derivation of the integrals for the entanglement between the coin degrees of freedom}\label{app:CoinEntOnLine}

    We now give a detailed derivation of the solution of the integral expression in \cref{eq:IntegralEqForPhi1}. 
    
    \subsection{Integral $I(n)$}

        For convenience, we write the integral, which appears in $I(n)$, as $i(n)$, i.e., $I(n) = (i(n)-1)/2$.
        Inserting \cref{eq:QWCoefficientPsi0,eq:QWCoefficientPsi1} and using the symmetries of the integrand of $i(n)$ to restrict the integration domain to $[0,\pi/2]$, we find
        \begin{align}
            i(n) ={}& \frac{2}{\pi} \int_{0}^{\pi/2} \mathrm{d}{k}\, \frac{1}{1+\cos^2k}  
            \\ \nonumber
            &{}- (-1)^n\frac{2}{\pi} \int_{0}^{\pi/2} \mathrm{d}{k}\, \frac{\cos(2n \arcsin(\sin k / \sqrt{2}))}{1+\cos^2 k} .
        \end{align}
        The first term is readily integrated using elementary methods,
        \begin{equation}
            \frac{2}{\pi} \int_{0}^{\pi/2} \mathrm{d}{k}\, \frac{1}{1+\cos^2k} = \frac{1}{\sqrt{2}}.
        \end{equation}
        The second integral can be rewritten using Chebyshev polynomials of the first kind, which are defined as $T_n(\cos\alpha) \equiv \cos(n\alpha)$.
        Applying the composition relation $T_{mn}(x) = T_m(T_n(x))$, we find
        \begin{equation}
            i(n) = \frac{1}{\sqrt{2}} - \frac{2}{\pi} (-1)^n \int_{0}^{\pi/2} \mathrm{d}{k}\, \frac{T_n(\cos^2 k)}{1+\cos^2k}.
        \end{equation}
        To solve the above integral, we construct the generating function $G_i(x) = \sum_{n=0}^\infty i(n) x^n$.
        Note that in the main text we used $I(n)$ to construct the generating function, giving a slight modification $G_i(x) = \frac{1}{2}(G(x)-1)$.
        We recognize the first term as a geometric series and the second term can be determined via the known generating function for the Chebyshev polynomials
        \begin{equation}
            \sum_{n=0}^{\infty} T_n(z) x^n = \frac{1- x z}{1-2xz+x^2} .
        \end{equation}
        Thus, we obtain
        \begin{align}
            G_i(x) &= \frac{1}{\sqrt{2}(1-x)} \nonumber
            \\
            &\quad - \frac{2}{\pi} \underbrace{\int_{0}^{\pi/2} \,\mathrm{d}{k}\, \frac{1+x \cos^2k}{(1+2x \cos^2k+x^2)(1+\cos^2k)}}_{h}.
        \end{align}
        This integral can be brought into a more convenient form by firstly substituting $u = \tan k$ and then performing a partial fraction decomposition.
        Using $\tan^2k+1=1/\cos^2 k$, we can rewrite the integral as
        \begin{align}
            h &= \int_{0}^{\pi/2} \,\frac{\mathrm{d}{k}}{\cos^2k}\, \frac{\tan^2k + 1 + x}{[2x+(1+x^2)(\tan^2k+1)](\tan^2k + 2)} \nonumber
            \\
            &= \int_{0}^{\infty}  \,\mathrm{d}{u}\, \frac{u^2+1+x}{[(1+x^2)u^2 + (1+x)^2](u^2+2)}.
        \end{align}
        A partial fraction decomposition with respect to $u$ then gives
        \begin{multline}
            h = \frac{1}{1-x}\int_{0}^{\infty} \,\mathrm{d}{u}\, \frac{1}{u^2+2}  
            \\
            - \frac{x(x+1)}{1-x} \int_{0}^{\infty} \,\mathrm{d}{u}\, \frac{1}{u^2(x^2+1)+(1+x)^2}.
        \end{multline}
        We use the following known integral formula
        \begin{equation}\label{eq:RationalIntegral}
            \int_0^\infty \frac{\mathrm{d}{u}}{a u^2+b} = \frac{\pi}{2 \sqrt{ab}}
        \end{equation}
        to find
        \begin{equation}
            h = \pi \left(  \frac{1}{2\sqrt{2}} \frac{1}{1-x} - \frac{1}{2} \frac{x}{(1-x)\sqrt{1+x^2}}  \right).
        \end{equation}
        This yields the final result for the generating function
        \begin{equation}
            G_i(x) = x\,\frac{1}{1-x}\,\frac{1}{\sqrt{1+x^2}}.
        \end{equation}
        We recognize the geometric series
        \begin{equation}
            \frac{1}{1-x} = \sum_{n=0}^\infty x^n
        \end{equation}
        and the binomial series
        \begin{equation}
            \frac{1}{\sqrt{1+x^2}} = \sum_{n=0}^\infty (-1)^n \binom{2n}{n} \frac{1}{4^n} x^{2n}
        \end{equation}
        and further note that the additional factor $x$ simply corresponds to an index shift.
        Combining both series, we arrive at the final result given in \cref{eq:SolutionIntegralIOfn}
        \begin{equation}
            i(n) = \sum_{k=0}^{\left\lfloor \frac{n-1}{2} \right\rfloor} (-1)^k \binom{2k}{k} \frac{1}{4^k}.
        \end{equation}

    \subsection{The integral $J(n)$} 
    
        In the next step, we can solve the integral $J(n)$ in \cref{eq:IntegralEqForPhi1}, which reads
        \begin{equation}
            J(n) = (-1)^n \int_{-\pi}^{\pi} \frac{\,\mathrm{d}{k}\,}{2\pi} \mathrm{Re}\left[ \rme^{-\rmi \phi} \tilde \psi^-_0(n,k) \tilde \psi^-_1(n,k) \right],
        \end{equation}
        by reducing this problem to the previously determined integral $I(n-1)$.
        First, we expand the exponential function and find the two terms 
        \begin{equation}
        \begin{aligned}
            {}& \cos\phi \, \mathrm{Re}\left[  \tilde \psi^-_0(n,k) \tilde \psi^-_1(n,k) \right]
            \\
            \text{and }
            {}& \sin\phi \, \mathrm{Re}\left[\rmi\,  \tilde \psi^-_0(n,k) \tilde \psi^-_1(n,k) \right].
        \end{aligned}
        \end{equation}
        Since $\tilde\psi^-_j(n,-k) = \tilde\psi^{-}_j(n,k)^*$ (for $j\in\{0,1\}$), we see
        \begin{equation}
            \mathrm{Re}\left[  \tilde \psi^-_0(n,-k) \tilde \psi^-_1(n,-k) \right] = \mathrm{Re}\left[  \tilde \psi^-_0(n,k) \tilde \psi^-_1(n,k) \right]
        \end{equation}
        and
        \begin{equation}
            \mathrm{Re}\left[\rmi\,  \tilde \psi^-_0(n,-k) \tilde \psi^-_1(n,-k) \right] = -\mathrm{Re}\left[\rmi\,  \tilde \psi^-_0(n,k) \tilde \psi^-_1(n,k) \right].
        \end{equation}
        Thus, the integral over the symmetric interval $k \in [-\pi,\pi]$ that contains the $\sin\phi$ term vanishes, and we are left with
        \begin{equation}
            J(n) = \cos\phi\, (-1)^n \int_{-\pi}^{\pi} \frac{\,\mathrm{d}{k}\,}{2\pi} \mathrm{Re}\left[ \tilde \psi^-_0(n,k) \tilde \psi^-_1(n,k) \right].
        \end{equation}
        We now substitute \cref{eq:QWCoefficientPsi0,eq:QWCoefficientPsi1} and find
        \begin{multline}
            \frac{1}{2} \int_{-\pi}^{\pi} \frac{\,\mathrm{d}{k}\, }{2\pi} \left[  - \frac{\cos^2 k}{1+\cos^2 k}  + (-1)^n \frac{\cos^2 k}{1+\cos^2 k} \cos( 2n\omega_k )\right.
            \\ \left. + (-1)^n \frac{\sin k}{\sqrt{1+\cos^2 k}} \sin( 2n\omega_k )  \right].
        \end{multline}
        To make the connection to $I(n)$, we consider
        \begin{equation}
            i(n-1) =\int_{-\pi}^{\pi} \frac{\,\mathrm{d}{k}\, }{2\pi} \frac{1+(-1)^n \cos( 2n\omega_k - 2\omega_k ) }{1+\cos^2 k}.
        \end{equation}
        Upon using the trigonometric addition formulas as well as
        \begin{equation}
        \begin{aligned}
            \cos(2n\omega_k) &= \cos^2k,
            \\
            \sin(2n\omega_k)&= \sqrt{1+\cos^2 k}\, \sin k,
        \end{aligned}
        \end{equation}
        we find
        \begin{equation}
            i(n-1) = 2 \frac{J(n)}{\cos\phi} + 1.
        \end{equation}
        This yields the final result
        \begin{equation}
            J(n) = \cos\phi \frac{i(n-1)-1}{2} = \cos\phi\, I(n-1).
        \end{equation}

    \subsection{Relation to the von Neumann entropy}
    
        The methodology developed here can be used to find the analytic expression of the von Neumann entropy $S_E(n)$ for the partition $\mathcal{H}_\text{coin} : \mathcal{H}_\text{position}$.
        This quantity is often considered, for example in Refs.\ \cite{carneiroEntanglementCoinedQuantum2005,abalQuantumWalkLine2006,ideEntanglementDiscretetimeQuantum2011,ortheyAsymptoticEntanglementQuantum2017}.
        However, either the asymptotic case $n \to \infty$ is considered or it is computed numerically only.
        Here, we provide an exact expression for all cases.
    
        For simplicity, we focus on the case of simple initial conditions $|\psi_0\rangle = |-,0\rangle$.
        In the notation of Ref.\ \cite{abalQuantumWalkLine2006}, the two integrals that appear are given by
        \begin{align}
            A(n) &= \frac{1}{2} \int_{-\pi}^{\pi} \frac{\,\mathrm{d}{k}\,}{2\pi} \frac{1-(-1)^n \cos(2 n \omega_k)}{1+\cos^2 k}
            \\
            B(n) &= \frac{\rmi}{2} \int_{-\pi}^{\pi} \frac{\,\mathrm{d}{k}\, }{2\pi} \frac{ \rme^{\rmi k}}{\sqrt{1+\cos^2 k}} \left[  \frac{\cos k}{\sqrt{1+\cos^2 k}}
            \right. \nonumber
            \\
             &\left.- (-1)^n \left(  \frac{\cos k}{\sqrt{1+\cos^2 k}} \cos( 2n\omega_k ) + \rmi \sin( 2n\omega_k )    \right)  \right] 
        \end{align}
        We note that there appears to be a typo in the integral for $B(n)$ that appears in Eq.\ (20) in Ref.\ \cite{abalQuantumWalkLine2006}. 
        The first term in their square brackets should read $\cos k / \sqrt{1+\cos^2 k}$ instead of $1$. 
        The asymptotic expression in Eq.\ (21), however, uses the correct term.
        These integrals can be solved straightforwardly using the methods presented above to yield
        \begin{align}
            A(n) &= \frac{1}{2} i(n)
            \\
            B(n) &= \rmi \frac{1-i(n+1)}{2}.
        \end{align}
        Thus, the eigenvalues from Eq.\ (16) in \cite{abalQuantumWalkLine2006} are given by
        \begin{align}
            & r_{1,2}(n)
            \\ \nonumber
            =& \frac{1}{2}\Big( 1 \pm \sqrt{2-i(n{+}1)[2-i(n{+}1)]-i(n)[2-i(n)]} \Big),
        \end{align}
        which yields the von Neumann entropy
        \begin{equation}
            S_E(n) = -[r_1(n) \log_2 r_1(n) + r_2(n) \log_2 r_2(n)] .
        \end{equation}

\bibliography{refs}

\end{document}